\documentclass[twocolumn,showpacs,preprintnumbers,amsmath,amssymb, prb]{revtex4}

\usepackage{graphicx}
\usepackage{dcolumn}
\usepackage{bm}

\begin{document}

\preprint{APS/123-QED}

\title{Antiferromagnetism in two-dimensional $t$-$J$ model: pseudospin representation}

\author{Daisuke Yamamoto}
\email{yamamoto@kh.phys.waseda.ac.jp}
\author{Susumu Kurihara}
\affiliation{Department of Physics, Waseda University, Okubo, Shinjuku, Tokyo 169-8555, Japan.}
\homepage{http://www.kh.phys.wasedas.ac.jp/}
\date{\today}

\begin{abstract}
We discuss a pseudospin representation of the two-dimensional $t$-$J$ model. We introduce pseudospins associated with empty sites, deriving a new representation of the $t$-$J$ model that consists of local spins and spinless fermions.
We show, within a mean-field approximation, that our representation of $t$-$J$ model corresponds to the {\it isotropic} antiferromagnetic Heisenberg model in an effective magnetic field. The strength and the direction of the effective field are determined by the hole doping ${\delta}$ and the orientation of pseudospins associated with empty sites, respectively. We find that the staggered magnetization in the standard representation corresponds to the component of magnetization perpendicular to the effective field in our pseudospin representation. Using a many-body Green's function method, we show that the staggered magnetization decreases with increasing hole doping ${\delta}$ and disappears at ${\delta \approx 0.06-0.15}$ for $t/J=2-5$. Our results are in good agreement with experiments and numerical calculations in contradistinction to usual mean-field methods.
\end{abstract}

\maketitle

\section{\label{Introduction}Introduction} 
It is well known that the essential physical properties of the copper oxide materials are described by the two-dimensional $t$-$J$ model~\cite{tJ1,tJ2,tJ3slave5}. The action of this model is restricted to the single occupancy sector of Hilbert space, $\sum_{\sigma}c_{i\sigma}^{\dagger}c_{i\sigma}\leq1$. At half-filling, the $t$-$J$ model reduces to an antiferromagnetic Heisenberg model. In the doped case, on the other hand, due to the single occupancy constraint, electrons can move only onto empty sites. Thus the constrained electron operators no longer obey the fermionic anti-commutation relations. 

The slave particle mean-field methods~\cite{slave1,slave3,slave4,tJ3slave5,slave6,slave7,slave8} were the first approaches to overcome this problem. In these methods, the constrained electron operators are expressed in terms of auxiliary fermions and bosons. However, as the local single occupancy constraints are replaced by an average global constraint, these mean-field methods lead to considerable errors: for example, the antiferromagnetic long-range order is overestimated~\cite{slave6,slave7,slave8}. 

Another approach~\cite{fs1,fs2,fs3,fs4,fs5} is to express the constrained electron operators in terms of spinless fermions and spin-$1/2$ operators by introducing pseudospins state associated with empty sites. Using this representation, the $t$-$J$ Hamiltonian can be described without any constraint in contrast with the slave particle methods. The spin-``up'' and spin-``down'' symmetry (time-reversal symmetry) of this representation is discussed by Wang and Rice~\cite{fs3} and Loos~\cite{fs4}. In the present paper, we develop this idea and calculate the critical doping ${\delta}_{c}$, where the antiferromagnetic long-range order disappears.

Our representation of the $t$-$J$ model corresponds to the {\it isotropic} antiferromagnetic Heisenberg model in an effective magnetic field within a mean-field approximation. In order to calculate the staggered magnetization of the model, we employ the many-body Green's function method developed by Fr\"obrich and Kuntz~\cite{mag7}. Unlike the slave particle methods, we show that the critical doping ${\delta}_{c}$ for the disappearance of antiferromagnetic long-range order is in good agreement with the numerical calculations\cite{nu1,nu2,nu3,nu4,nu5}. 

This paper is organized as follows. In Sec.~\ref{2}, we introduce a transformation mapping the original Hilbert space of the constrained electrons to the tensor-product space of the spinless fermion and spin states. We also discuss the difference between our method and the preceding methods~\cite{fs3,fs4}. In Sec.~\ref{3}, using the many-body Green's function method, we calculate the staggered magnetization of the model obtained in Sec.~\ref{2} within a mean-field approximation. A summary is presented in Sec.~\ref{4}.

\section{\label{2}Pseudospin representation}

We consider the $t$-$J$ Hamiltonian
\begin{eqnarray}
H=-t\!\!\sum_{<i,j>\sigma}\!\!(c_{i\sigma}^\dagger
c_{j\sigma}+{\rm h.c.})-\mu\sum_{i\sigma}\!c_{i\sigma}^\dagger
c_{i\sigma}+J\!\!\sum_{<i,j>}\!\!\mathbf{s}_i\cdot \mathbf{s}_j, 
\label{HtJ}
\end{eqnarray}
where $c_{i\sigma}$ is the electron annihilation operator with spin $\sigma$ at site $i$, $\mu$ is the chemical potential, and $\mathbf{s}_{i}=\frac{1}{2}\sum_{\alpha\beta}c_{i\alpha}^{\dagger}\mathbf{\tau}_{\alpha\beta}c_{i\beta}$ is the spin operator with the Pauli matrices $\mathbf{\tau}$. The sum is taken over all nearest neighbor bonds. 
The electron in this model is subjected to the constraint of no double occupancy, {\it i.e.},
\begin{equation}
    \sum_{\sigma}c_{i\sigma}^{\dagger}c_{i\sigma}\leq1.
\label{constraint}
\end{equation}
Consequently, the basis of possible states in the $t$-$J$ model consist of the states $|i\sigma\rangle$ and $|i0\rangle$, which correspond to a site singly occupied by an electron with spin $\sigma$ and to an empty site, respectively. The constrained electron operators $c_{i\sigma}$ act on this basis as
\begin{eqnarray}
c_{i\sigma}|i\sigma\rangle=|i0\rangle, \ \ \ c_{i\sigma}^{\dagger}|i0\rangle=|i\sigma\rangle.
\label{action}
\end{eqnarray}
\subsection{Pseudospin associated with empty site}
One of the most popular techniques to handle the single occupancy constraint~(\ref{constraint}) is the slave particle method~\cite{slave1,slave3,slave4,tJ3slave5,slave6,slave7,slave8}. In the slave boson representation, the constrained electron operators are given by the mapping $c_{i\sigma}\rightarrow b_{i}^{\dagger}f_{i\sigma}$, where $b_{i}$ is the bosonic operator annihilating the empty state and $f_{i\sigma}$ is the fermionic operator annihilating the single occupied state. In this case, the non-holonomic constraint~(\ref{constraint}) is replaced by the holonomic constraint $b_{i}^{\dagger}b_{i}+\sum_{\sigma}f_{i\sigma}^{\dagger}f_{i\sigma}=1$. For practical purposes, however, these local constraints for each site are almost inevitably replaced by an average global constraint, resulting in errors: one of the most serious drawbacks is that the antiferromagnetic long-range order is overestimated~\cite{slave6,slave7,slave8}.

In this paper, we develop another approach, rewriting the $t$-$J$ model in terms of spinless fermionic operators (charge degree of freedom) and local spin-$\frac{1}{2}$ operators~\cite{fs1,fs2,fs3,fs4,fs5}. In this approach, the Hilbert space is mapped onto a tensor product space, $|\cdot\rangle_{h}\otimes|\cdot\rangle_{S}$, as 
\begin{eqnarray}
|i\sigma\rangle \rightarrow |i0\rangle_{h}|i\sigma\rangle_{S},~~~ 
|i0\rangle \rightarrow |i1\rangle_{h}|iS\rangle_{S}, 
\label{state}
\end{eqnarray}
where
\begin{eqnarray}
|iS\rangle_{S}\!\equiv \!C_{1}|i\!\uparrow\rangle_{S}+C_{2}|i\!\downarrow\rangle_{S},
\label{pseudospin}
\end{eqnarray}
with $|C_{1}|^{2}+|C_{2}|^{2}=1$, is the pseudospin state associated with an empty site. We introduce the fermionic ``holon'' operator $h_{i}$ and the ``spin'' operator $S_{i}$ acting as
\begin{eqnarray}
h_{i}^{\dagger}|i0\rangle_{h}\!\!&=&\!\!|i1\rangle_{h}, ~~~~h_{i}|i1\rangle_{h}=|i0\rangle_{h}, \nonumber \\
S_{i}^{+}|i\!\downarrow\rangle_{S}\!\!&=&\!\!|i\!\uparrow\rangle_{S}, ~~S_{i}^{-}|i\!\uparrow\rangle_{S}=|i\!\downarrow\rangle_{S}.
\label{hs}
\end{eqnarray}
Owing to Eq.~(\ref{state}), the constrained electron operators are written as
\begin{eqnarray}
c_{i\uparrow}\rightarrow \tilde{c}_{i\uparrow}\!\!&=&\!\!h_{i}^{\dagger}(C_{1}S_{i}^{+}S_{i}^{-}+C_{2}S_{i}^{-}), \nonumber \\
c_{i\downarrow}\rightarrow \tilde{c}_{i\downarrow}\!\!&=&\!\!h_{i}^{\dagger}(C_{1}S_{i}^{+}+C_{2}S_{i}^{-}S_{i}^{+}).
\label{transform}
\end{eqnarray}
Clearly, the new operator $\tilde{c}_{i\sigma}$ has the same action as the operator $c_{i\sigma}$ on the basis vectors~(\ref{state}). The four-dimensional space $|\cdot\rangle_{h}\otimes|\cdot\rangle_{S}$ consists of the orthonormal vectors $|i0\rangle_{h}|i\sigma\rangle_{S}, |i1\rangle_{h}|iS\rangle_{S}, $ and $|i1\rangle_{h}|i\bar{S}\rangle_{S}$ with $|i\bar{S}\rangle_{S}\!\equiv \!C_{1}^{*}|i\!\downarrow\rangle_{S}-C_{2}^{*}|i\!\uparrow\rangle_{S}$. However, it is easy to prove that the operator $\tilde{c}_{i\sigma}$ acts on the $|i1\rangle_{h}|i\bar{S}\rangle_{S}$ state as
\begin{eqnarray}
\tilde{c}_{i\sigma}|i1\rangle_{h}|i\bar{S}\rangle_{S}=0,~~\tilde{c}^{\dagger}_{i\sigma}|i1\rangle_{h}|i\bar{S}\rangle_{S}=0.
\label{SS}
\end{eqnarray}
Consequently, the unphysical basis vector $|i1\rangle_{h}|i\bar{S}\rangle_{S}$ can be ignored when the action of $\tilde{c}_{i\sigma}$ is considered~\cite{fs4}.

In this pseudospin representation, the ``true'' spin operator and the constrained electron number operator are expressed as
\begin{eqnarray}
\mathbf{s}_{i}&\rightarrow&(1-h_{i}^{\dagger}h_{i})\mathbf{S}_{i}.
\label{spintrans}, \\
\sum_{\sigma}c_{i\sigma}^{\dagger}c_{i\sigma}&\rightarrow&\sum_{\sigma}\tilde{c}_{i\sigma}^{\dagger}\tilde{c}_{i\sigma}=1-h_{i}^{\dagger}h_{i}  \label{number}.
\end{eqnarray}
Consequently, the single occupancy constraint
\begin{eqnarray}
\sum_{\sigma}c_{i\sigma}^{\dagger}c_{i\sigma}\leq1\rightarrow 1-h_{i}^{\dagger}h_{i}\leq1
\label{constraint2}
\end{eqnarray}
is automatically satisfied. Thus no approximations for the constraint are necessary unlike the slave particle methods~\cite{slave1,slave3,slave4,tJ3slave5,slave6,slave7,slave8}. 
\subsection{Pseudospin representation of $t$-$J$ model}
In our pseudospin representation, the $t$-$J$ model is found to be expressed as
\begin{eqnarray}
\tilde{H}=-t\!\!\sum_{<i,j>}\!\!\Bigl[h_{i}h_{j}^{\dagger}\bigl\{C_{1}^{2}(S_{i}^{+}S_{i}^{-}S_{j}^{+}S_{j}^{-}+S_{i}^{-}S_{j}^{+})\qquad \qquad \nonumber \\+C_{1}C_{2}^{*}(S_{i}^{+}S_{j}^{+}S_{j}^{-}+S_{i}^{-}S_{i}^{+}S_{j}^{+})\qquad \qquad \nonumber \\+C_{1}^{*}C_{2}(S_{i}^{+}S_{i}^{-}S_{j}^{-}+S_{i}^{-}S_{j}^{-}S_{j}^{+})\qquad \qquad \nonumber \\  +C_{2}^{2}(S_{i}^{-}S_{i}^{+}S_{j}^{-}S_{j}^{+}+S_{i}^{+}S_{j}^{-})\bigl\}+{\rm h.c.}\Bigl] \qquad \nonumber \\
-\mu\sum_{i}\!h_{i}^\dagger h_{i}
+J\!\!\sum_{<i,j>}\!\!h_{i}h_{i}^{\dagger}\mathbf{S}_i\cdot \mathbf{S}_{j}h_{j}h_{j}^{\dagger},\qquad 
\label{pr}
\end{eqnarray}
which can be rewritten in the following more convenient form with $C_{1}=\cos\frac{\theta}{2}$ and $C_{2}=e^{i\varphi}\sin\frac{\theta}{2}$ [Fig.~\ref{fig:orientation}]:
\begin{figure}[t]
\begin{center}
\includegraphics[height=55mm]{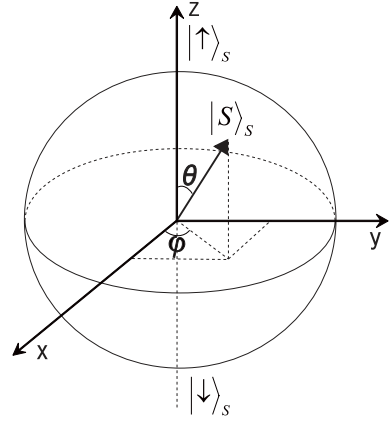}
\caption{\label{fig:orientation} Bloch sphere describing the orientation of pseudospins.}
\end{center}
\end{figure}
\begin{eqnarray}
 \tilde{H}=-t\!\!\sum_{<i,j>}\!\!\!\!&&\!\!\left[h_{i}h_{j}^{\dagger}\left\{ \mathbf{S}_i\cdot \mathbf{S}_{j}+\frac{1}{4} \right. \right. \nonumber \\ 
 &&+\left. \left.  \mathbf{n}\cdot\left(\frac{1}{2}\mathbf{S}_{i}+\frac{1}{2}\mathbf{S}_{j}+i\mathbf{S}_{i}\times\mathbf{S}_{j}\right)   \right\}+{\rm h.c.}\right]\nonumber \\
 &&-\mu\sum_{i}\!h_{i}^\dagger h_{i}+J\!\!\sum_{<i,j>}\!\!h_{i}h_{i}^{\dagger}\mathbf{S}_i\cdot \mathbf{S}_{j}h_{j}h_{j}^{\dagger},\qquad 
\label{pr2}
\end{eqnarray}
where $\mathbf{n}=(\sin\theta\cos\varphi,\sin\theta\sin\varphi,\cos\theta)$ is a unit vector in the orientation of pseudospins. 
Loos~\cite{fs4} put a certain restriction on the orientation of the pseudospins so that the time-reversal symmetry of the $t$-$J$ model is preserved, assuming that the ``spin'' operators $\mathbf{S}_{i}$ are transformed in the time-reversal operation similarly to the ``true'' spin operators $\mathbf{s}_{i}$.
In contrast, we put no restriction on the parameters $\theta$ and $\varphi$. In Sec.~\ref{3B}, we show that our results are independent of the parameters $\theta$ and $\varphi$.

The representation of the $t$-$J$ model introduced in Ref.~\onlinecite{fs2} can be obtained by setting $\theta=0$ in Eq.~(\ref{pr2}). The similar expression is also introduced for the Hubbard model in Ref.~\onlinecite{fs5}. For simplicity, in this paper, we also choose the parameters as $\theta=0$, {\it i.e.}, the pseudospins are aligned along the $z$ axis. Consequently, the pseudospin representation of the $t$-$J$ model is obtained as
\begin{eqnarray}
 \tilde{H}=-t\!\!\sum_{<i,j>}\!\!\!\!&&\!\!\left[h_{i}h_{j}^{\dagger}\left\{ \mathbf{S}_i\cdot \mathbf{S}_{j}+\frac{1}{4} \right. \right. \nonumber \\ 
 &&+\left. \left. \frac{1}{2}S_{i}^{z}+\frac{1}{2}S_{j}^{z}+i(\mathbf{S}_{i}\times\mathbf{S}_{j})_{z}  \right\}+{\rm h.c.}\right]\nonumber \\
 &&-\mu\sum_{i}\!h_{i}^\dagger h_{i}+J\!\!\sum_{<i,j>}\!\!h_{i}h_{i}^{\dagger}\mathbf{S}_i\cdot \mathbf{S}_{j}h_{j}h_{j}^{\dagger}.\qquad 
\label{hamiltonian}
\end{eqnarray}
Only for ``true'' spins, the $J$-term in Eq.~(\ref{hamiltonian}) represents the ordinary magnetic energy. At half-filling, the $t$-$J$ Hamiltonian $\tilde{H}$ [Eq.~(\ref{hamiltonian})] is reduced to the Heisenberg Hamiltonian. 
\subsection{Mean-field approximation}
Within the mean-field approximation, the $t$-$J$ Hamiltonian $\tilde{H}$ [Eq.~(\ref{hamiltonian})] can be decoupled into a holon part, $\tilde{H}_{\rm h}^{{\rm MF}}$, and a spin part, $\tilde{H}_{\rm S}^{{\rm MF}}$, as
\begin{eqnarray}
\tilde{H}_{\rm h}^{{\rm MF}}\!&=&\!t_{\rm eff}\!\!\sum_{<i,j>}\!\!  (h_{j}^{\dagger}h_{i}+{\rm h.c.})-\mu\sum_{i}\!h_{i}^\dagger h_{i},\label{charge}
 \\
\tilde{H}_{\rm S}^{{\rm MF}}\!&=&\!J_{\rm eff}\!\!\sum_{<i,j>}\!\!\mathbf{S}_i\cdot \mathbf{S}_{j}-B_{\rm eff}^{z}\!\!\sum_{i}\!{S_{i}^{z}}
\label{spin}
\end{eqnarray}
with $t_{\rm eff}=t\langle \mathbf{S}_i\cdot \mathbf{S}_{j}+\frac{1}{4}+\frac{1}{2}S_{i}^{z}+\frac{1}{2}S_{j}^{z}
 \rangle$, $J_{\rm eff}=J\{(1-\delta)^{2}-\phi^{2}\}+2t\phi$, and $B_{\rm eff}^{z}=-4t\phi$, where $\langle \cdots \rangle$ is an average over the ensemble. The hole doping and the particle-hole order parameter are defined as $\delta=\langle h_{i}^{\dagger}h_{i} \rangle$ and $\phi=\langle h_{i}^{\dagger}h_{i+\eta} \rangle$ with $\eta=\pm\hat{x},\pm\hat{y}$, respectively. 

The spin part $\tilde{H}_{\rm S}^{{\rm MF}}$ [Eq.~(\ref{spin})] corresponds to the Hamiltonian describing the {\it isotropic} antiferromagnet in a uniform magnetic field with the effective exchange constant $J_{\rm eff}$ and the effective magnetic field $B_{\rm eff}^{z}$ along $z$ axis. We note that the direction of the effective field depends on the choice of the orientation of the pseudospin. 
\section{\label{3}Doping dependence of staggered magnetization}
In this section, we consider the doping dependence of the staggered magnetization of the mean-field Hamiltonian~(\ref{charge})~and~(\ref{spin}). In Sec.~\ref{3A}, we calculate the components of magnetization for the Heisenberg antiferromagnet
\begin{eqnarray}
\bar{H}=\sum_{<i,j>}\!\!(\bar{J}\mathbf{S}_i\cdot \mathbf{S}_{j}+D^{x}S_{i}^{x}S_{j}^{x})-B^{z}\!\!\sum_{i}\!{S_{i}^{z}},
\label{Heisenberg}
\end{eqnarray}
where $\bar{J}$ is the exchange constant, $D^{x}$ is the anisotropy parameter, and $B^{z}$ is the magnetic field along $z$ axis. The Heisenberg model~(\ref{Heisenberg}) corresponds to the spin part of the mean-field Hamiltonian~(\ref{spin}) with
\begin{eqnarray}
\bar{J}\rightarrow J_{\rm eff},~~~D^{x}\rightarrow 0,~~~B^{z}\rightarrow B_{\rm eff}^{z}.
\label{correspondence}
\end{eqnarray}
Using this correspondence, we calculate the staggered magnetization of the mean-field Hamiltonian (\ref{charge}) and (\ref{spin}) in Sec.~\ref{3B}.

\subsection{\label{3A}Antiferromagnet in a magnetic field}
The magnetic properties of two-dimensional antiferromagnets have been the focus of many preceding theoretical studies~\cite{mag1,mag2,mag3,mag4,mag5,mag6,mag7}. In particular, the properties of a two-dimensional anisotropic antiferromagnet in a transverse field were recently studied with rotating frame method~\cite{mag6} and in the non-rotating frame~\cite{mag7}, of which difference were discussed by Fr\"obrich and Kuntz~\cite{mag7}. In this subsection, we calculate the components of magnetization of the Heisenberg model~(\ref{Heisenberg}) directly in the non-rotating frame as in Ref.~\onlinecite{mag7}.
The retarded Green's functions are defined as
\begin{eqnarray}
G_{ij}^{\alpha -}(t-t')\!&=&\!-i\theta(t-t')\langle [S_{i}^{\alpha}(t),S_{j}^{-}(t')] \rangle \nonumber \\
\!&=&\!\langle\langle S_{i}^{\alpha};S_{j}^{-}\rangle\rangle~;~~\alpha=+,-,z~~,
\label{Green}
\end{eqnarray}
where $\theta(t)$ is the step function. The Fourier transform of the Green's functions are denoted by $\langle\langle S_{i}^{\alpha};S_{j}^{-}\rangle\rangle_{\omega}$ in the energy space. The equations of motion for the Green's functions given as Eq.(\ref{Green}) are expressed as
\begin{eqnarray}
\omega G_{ij}^{\alpha -}(\omega)=\left(
\begin{array}{c} 2\langle S_{i}^{z} \rangle\delta_{ij} \\
0 \\
-\langle S_{i}^{x} \rangle\delta_{ij}
 \end{array} \right)+\langle\langle[S_{i}^{\alpha},\bar{H}];S_{j}^{-}\rangle\rangle_{\omega}.
\label{Eqofmotion}
\end{eqnarray}
In order to close the system of equations, we adopt the Tyablikov (RPA)-decoupling~\cite{tyablikov} of higher-order Green's functions;
\begin{eqnarray}
\langle\langle S_{i}^{\alpha}S_{k}^{\beta};S_{j}^{-}\rangle\rangle_{\omega}\approx\langle S_{i}^{\alpha}\rangle G_{kj}^{\beta -}+\langle S_{k}^{\beta}\rangle G_{ij}^{\alpha -}.
\label{RPA}
\end{eqnarray}
Fourier transformations to momentum space are given by
\begin{eqnarray}
G_{\mu\nu}^{\alpha -}({\bf q})=\frac{2}{N}\sum_{i_{\mu}j_{\nu}}G_{i_{\mu}j_{\nu}}^{\alpha -}e^{-i{\bf q}\cdot({\bf R}_{i_{\mu}}-{\bf R}_{j_{\nu}})}, 
\label{Fourier}
\end{eqnarray}
where subscripts $\mu$, $\nu$ denote sublattice indices (A or B). Each sublattice consists of $N/2$ lattice sites. Furthermore, to simplify the calculations, we assume that the magnetic components can be defined as $\langle S_{i_{\rm A}}^{x}\rangle=-\langle S_{i_{\rm B}}^{x}\rangle\equiv m^{x}$ and $\langle S_{i_{\rm A}}^{z}\rangle=\langle S_{i_{\rm B}}^{z}\rangle\equiv m^{z}$ due to the symmetry of the present case. We now rewrite Eq.(\ref{Eqofmotion}) in a matrix form,
\begin{eqnarray}
\left(\omega {\bf 1}-\left(
\begin{array}{c c} \mbox{\boldmath $\Gamma$} & 0 \\
0 & \mbox{\boldmath $\Gamma$}
\end{array} \right) \right) \left(
\begin{array}{c} {\bf G}_{\rm A} \\
{\bf G}_{\rm B}
\end{array} \right)=\left(
\begin{array}{c} \mbox{\boldmath $\Lambda$}_{\rm A} \\ 
\mbox{\boldmath $\Lambda$}_{\rm B}
\end{array}
\right).
\label{mateq}
\end{eqnarray}
In Eq.(\ref{mateq}), the diagonal $6\times 6$ sub-matrix $\mbox{\boldmath $\Gamma$}$ is given by
\begin{eqnarray}
\mbox{\boldmath $\Gamma$}=\left(\begin{array}{cccccc}
-a&0&-b&c_{\bf q}&d_{\bf q}&-e_{\bf q}\\
0&a&b&-d_{\bf q}&-c_{\bf q}&e_{\bf q}\\
-\frac{1}{2}b&\frac{1}{2}b&0&-\frac{1}{2}e_{\bf q}&\frac{1}{2}e_{\bf q}&0\\
c_{\bf q}&d_{\bf q}&e_{\bf q}&-a&0&b\\
-d_{\bf q}&-c_{\bf q}&-e_{\bf q}&0&a&-b\\
\frac{1}{2}e_{\bf q}&-\frac{1}{2}e_{\bf q}&0&\frac{1}{2}b&-\frac{1}{2}b&0
\end{array}\right)
,
\label{gamma}
\end{eqnarray}
with $a=zJm^{z}+B^{z}$, $b=z(J+D^{x})m^{x}$, $c_{\bf q}=z(J+\frac{1}{2}D^{x})\gamma_{\bf q}m^{z}$, $d_{\bf q}=\frac{1}{2}zD^{x}\gamma_{\bf q}m^{z}$, and $e_{\bf q}=zJ\gamma_{\bf q}m^{x}$, where $z=4$ is the number of nearest neighbors and $\gamma_{\bf q}=\frac{1}{2}(\cos q_{x}+\cos q_{y})$ is the Fourier factor for a square lattice: we set the lattice constant to be unity. The sublattice Green's functions ${\bf G}_{\mu}$ are defined as
\begin{eqnarray}
{\bf G}_{\mu}=\left(\begin{array}{c}
{\bf G}_{{\rm A}\mu}\\
{\bf G}_{{\rm B}\mu}
\end{array}\right)
;~~\mu={\rm A},{\rm B}~,
\label{GG}
\end{eqnarray}
with
\begin{eqnarray}
{\bf G}_{\mu\nu}=\left(\begin{array}{c}
G_{\mu\nu}^{+-}\\
G_{\mu\nu}^{--}\\
G_{\mu\nu}^{z-}
\end{array}\right);~~\mu,\nu={\rm A},{\rm B}~,
\label{Gmat}
\end{eqnarray}
while the vectors $\mbox{\boldmath $\Lambda$}_{\mu}$ are given by
\begin{eqnarray}
\mbox{\boldmath $\Lambda$}_{\mu}=\left(\begin{array}{c}
\mbox{\boldmath $\Lambda$}_{{\rm A}\mu}\delta_{{\rm A}\mu}\\
\mbox{\boldmath $\Lambda$}_{{\rm B}\mu}\delta_{{\rm B}\mu}
\end{array}\right)
;~~\mu={\rm A},{\rm B}~,
\label{LL}
\end{eqnarray}
with
\begin{eqnarray}
\mbox{\boldmath $\Lambda$}_{\mu\nu}=\left(\begin{array}{c}
2\langle S_{i_{\nu}}^{z}\rangle\\
0\\
-\langle S_{i_\nu}^{x} \rangle
\end{array}\right);~~\mu,\nu={\rm A},{\rm B}~.
\label{Lmat}
\end{eqnarray}
By solving Eq.~(\ref{mateq}), we obtain the Green's functions with the poles
\begin{eqnarray}
\omega\!&=&\!0,~~\pm \omega_{1},~~\pm \omega_{2} ~~,\nonumber \\
\omega_{1}\!&=&\!\sqrt{(a+c)^{2}+b^2 -d^2 -e^2},\nonumber \\
\omega_{2}\!&=&\!\sqrt{(a-c)^{2}+b^2 -d^2 -e^2}.
\label{poles}
\end{eqnarray}
Dealing with the pole $\omega=0$ is known to be difficult, in that the anticommutator Green's function is required~\cite{tyablikov2}. However, we can avoid this difficulty by introducing the new expression
\begin{eqnarray}
G_{\rm AA}^{+-}\!-\!G_{\rm AA}^{--}\!=\!\frac{1}{2\omega_{1}}\!\Bigl( \frac{m^{x}(b+e)-m^{z}(a+c+d)+m^{z}\omega_{1}}{\omega-\omega_{1}} \nonumber \\
-\frac{m^{x}(b+e)-m^{z}(a+c+d)-m^{z}\omega_{1}}{\omega+\omega_{1}} \Bigl)\nonumber \\
+\frac{1}{2\omega_{2}}\!\Bigl( \frac{m^{x}(b-e)-m^{z}(a-c-d)+m^{z}\omega_{2}}{\omega-\omega_{2}} \nonumber \\
-\frac{m^{x}(b-e)-m^{z}(a-c-d)-m^{z}\omega_{2}}{\omega+\omega_{2}} \Bigl).\nonumber \\
\label{GMG}
\end{eqnarray}
Since the pole $\omega=0$ is absent in this expression, we can use the standard spectral theorem~\cite{tyablikov2}.
Applying this theorem to Eq.(\ref{GMG}), we derive
\begin{eqnarray}
\frac{1}{2}=\frac{1}{N}\!\!\sum_{\bf q}\!\biggl[\frac{m^{x}(b+e)-m^{z}(a+c+d)}{2\omega_{1}}\coth \frac{\beta \omega_{1}}{2} \nonumber \\
+\frac{m^{x}(b-e)-m^{z}(a-c-d)}{2\omega_{2}}\coth \frac{\beta \omega_{2}}{2}\biggl].
\label{selfz}
\end{eqnarray}
In Eq.(\ref{selfz}), the sum runs over the first Brillouin zone. Moreover, using the fact that the commutator Green's function has to be regular at the origin~\cite{mag7}, {\it i.e.},
\begin{eqnarray}
\lim_{\omega \rightarrow 0}\omega G_{ij}^{z-}=0~~~,
\label{regularity}
\end{eqnarray}
we obtain
\begin{eqnarray}
m^{x}=\frac{1}{N}\!\!\sum_{\bf q}\!\biggl[\frac{(b-e)\{m^{x}(b+e)-m^{z}(a+c+d)\}}{2\omega_{1}^{2}} \nonumber \\
+\frac{(b+e)\{m^{x}(b-e)-m^{z}(a-c-d)\}}{2\omega_{2}^2}\biggl].
\label{selfx}
\end{eqnarray}
\begin{figure}[t]
\begin{center}
\includegraphics[height=53mm]{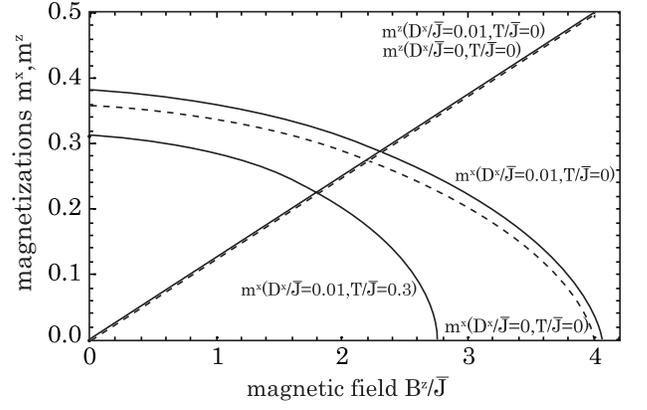}
\caption{\label{fig:field} Components of magnetization $m^x$, $m^z$ as function of $B^z$ at different temperatures $T$ for $D^x/\bar{J}=0.01$ (Solid lines) and $D^x/\bar{J}=0$ (dotted lines).}
\end{center}
\end{figure}
From Eqs.~(\ref{selfz})~and~(\ref{selfx}), all components of the magnetization, $m^{x}$ and $m^{z}$, are determined. As seen in the results in Fig.~\ref{fig:field}, the magnetization $m^{x}$ perpendicular to the magnetic field decreases with increasing the field strength, and disappears around field strength $B^{z}/\bar{J} \approx 4$ at $T=0$ for $D^{x}/\bar{J} =0$.

\subsection{\label{3B}Doping dependence of staggered magnetization}
In this subsection, we consider the staggered magnetization of the mean-field Hamiltonian~(\ref{charge})~and~(\ref{spin}). From the transformation~(\ref{spintrans}), the staggered magnetization in the original representation $|\langle s_{i}^{x}\rangle|$ is transformed into the pseudospin representation as
\begin{eqnarray}
|\langle s_{i}^{x}\rangle |\rightarrow |\langle (1-h_{i}^{\dagger}h_{i})S_{i}^{x}\rangle |\!\!&=&\!\!|\langle S_{i}^{x}\rangle |-|\langle h_{i}^{\dagger}h_{i}S_{i}^{x}\rangle | \nonumber \\
\!\!&=&\!\!|\langle S_{i}^{x}\rangle |\nonumber  \\
\!\!&\equiv&\!\! M.
\label{mag}
\end{eqnarray}
Equation~(\ref{mag}) shows that the staggered magnetization can be regarded as the component of magnetization perpendicular to the effective magnetic field (to the orientation of pseudospin) in the pseudospin representation. When deriving the relation~(\ref{mag}), we use the fact $\langle h_{i}^{\dagger}h_{i}S_{i}^{x}\rangle = 0$, choosing the orientation of pseudospin along $z$ axis. Thus the other components of magnetization cannot fulfill the similar relation to Eq.~(\ref{mag}). For example, one can easily show that $|\langle s_{i}^{z}\rangle |\rightarrow\hspace{-1.1em}/ ~~|\langle S_{i}^{z}\rangle |$.

\begin{figure}[t]
\begin{center}
\includegraphics[height=95mm]{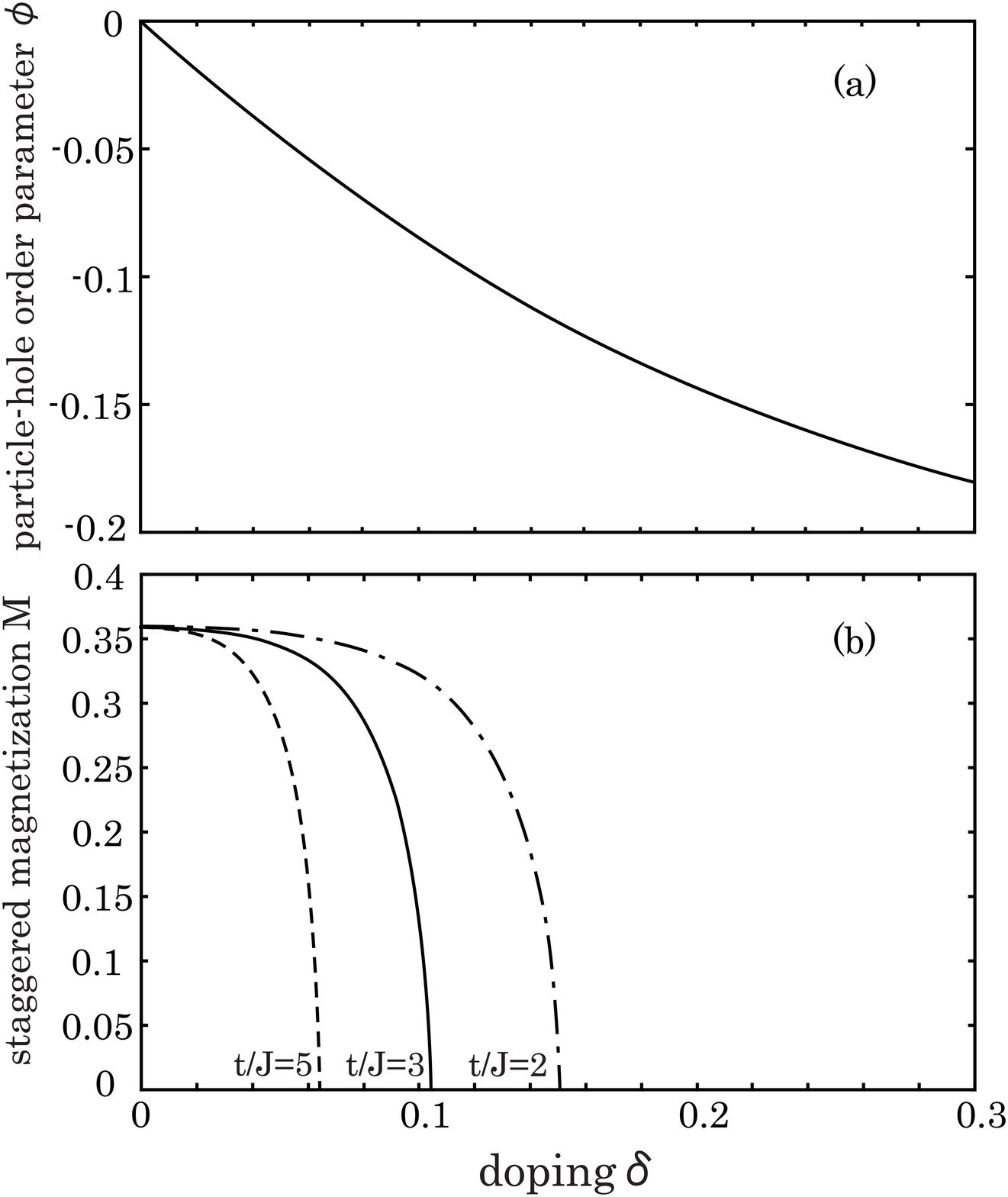}
\caption{\label{fig:dope} (a)~Particle-hole order parameter $\phi$ as function of doping $\delta$ at $T=0$ (independent of $t/J$). (b)~Staggered magnetization $M$ as function of doping at $T=0$ for $t/J=2$, $t/J=3$, and $t/J=5$. }
\label{fig:doping}
\end{center}
\end{figure}

Instead of starting from the Heisenberg model~(\ref{Heisenberg}), we now examine the spin part of the mean-field Hamiltonian~(\ref{spin}). After some algebra similar to the one given in Sec.~\ref{3A}, we obtain
\begin{eqnarray}
\frac{1}{2}=\frac{1}{N}\!\!\sum_{\bf q}\!\biggl[\frac{M(\tilde{b}+\tilde{e})-\tilde{M}(\tilde{a}+\tilde{c}+\tilde{d})}{2\tilde{\omega}_{1}}\coth \frac{\beta \tilde{\omega}_{1}}{2} \nonumber \\
+\frac{M(\tilde{b}-\tilde{e})-\tilde{M}(\tilde{a}-\tilde{c}-\tilde{d})}{2\tilde{\omega}_{2}}\coth \frac{\beta \tilde{\omega}_{2}}{2}\biggl], \label{selfM} \\
M=\frac{1}{N}\!\!\sum_{\bf q}\!\biggl[\frac{(\tilde{b}-\tilde{e})\{M(\tilde{b}+\tilde{e})-\tilde{M}(\tilde{a}+\tilde{c}+\tilde{d})\}}{2\tilde{\omega}_{1}^{2}} \nonumber \\
+\frac{(b+e)\{M(\tilde{b}-\tilde{e})-\tilde{M}(\tilde{a}-\tilde{c}-\tilde{d})\}}{2\tilde{\omega}_{2}^2}\biggl],
\label{selftM}
\end{eqnarray}
\begin{eqnarray}
\tilde{\omega}_{1}\!\!&=&\!\!\sqrt{(\tilde{a}+\tilde{c})^{2}+\tilde{b}^2 -\tilde{d}^2 -\tilde{e}^2}, \\
\tilde{\omega}_{2}\!\!&=&\!\!\sqrt{(\tilde{a}-\tilde{c})^{2}+\tilde{b}^2 -\tilde{d}^2 -\tilde{e}^2},
\label{tildeomega}
\end{eqnarray}
with $\tilde{M}=|\langle S_{i}^{z}\rangle|$, $\tilde{a}=zJ_{\rm eff}\tilde{M}+B_{\rm eff}^{z}$, $\tilde{b}=zJ_{\rm eff}M$, $\tilde{c}_{\bf q}=zJ_{\rm eff}\gamma_{\bf q}\tilde{M}$, $\tilde{d}_{\bf q}=0$, and $\tilde{e}_{\bf q}=zJ_{\rm eff}\gamma_{\bf q}M$. In addition to Eqs.~(\ref{selfM})~and~(\ref{selftM}), the equations for the holon Green's functions are needed to obtain a closed system of equations. The holon Green's function are defined as
\begin{eqnarray}
g_{ij}(t-t')\!&=&\!-i\theta(t-t')\langle \{h_{i}(t),h_{j}^{\dagger}(t')\} \rangle \nonumber \\
\!&=&\!\langle\langle h_{i};h_{j}^{\dagger}\rangle\rangle.
\label{hGreen}
\end{eqnarray}
From the holon part of the mean-field Hamiltonian~(\ref{charge}), we obtain the following expression of the holon Green's function;
\begin{eqnarray}
g_{{\bf q}}(\omega)=\frac{1}{\omega-(\varepsilon_{\bf q}-\mu)},
\label{hGreenq}
\end{eqnarray}
with $\varepsilon_{\bf q}=zt_{\rm eff}\gamma_{\bf q}$.
Using the spectral theorem~\cite{tyablikov2}, we also find
\begin{eqnarray}
\delta\!\!&=&\!\!\frac{1}{2N}\!\!\sum_{\bf q}\!\biggl[1-\tanh \frac{\beta(\varepsilon_{\bf q}-\mu)}{2}\biggl], \label{selfmu} \\
\phi\!\!&=&\!\!\frac{1}{2N}\!\!\sum_{\bf q}\!\gamma_{\bf q}\biggl[1-\tanh \frac{\beta(\varepsilon_{\bf q}-\mu)}{2}\biggl].
\label{selffai}
\end{eqnarray}
The set of self-consistency equations (\ref{selfM}), (\ref{selftM}), (\ref{selfmu}), and (\ref{selffai}) are numerically solved. In Figure~\ref{fig:doping}, we show the staggered magnetization and the particle-hole order parameter at $T=0$ as functions of doping $\delta$. 
The staggered magnetization decreases with increasing $\delta$, and disappears at the critical doping $\delta_{c}$, which is calculated to be $\delta_{c}\approx 0.105$ for $t/J=3$. In usual mean-field theories, since the single occupancy constraint~(\ref{constraint}) is treated only on the average, the antiferromagnetic order is generally overestimated. However, our results derived without the constraint are in good accord with the numerical results in Ref.~\onlinecite{nu1,nu2,nu3,nu4,nu5}. 

In this paper, we choose the pseudospin state as $|i\!\uparrow\rangle_{S}$. If the pseudospin state is chosen as $(|i\!\uparrow\rangle_{S}+|i\!\downarrow\rangle_{S})/2$, the direction of the effective magnetic field is aligned along the $x$ axis, and the staggered magnetization is mapped to the component of magnetization $|\langle S_{i}^{z}\rangle |$ (perpendicular to the effective field). 

\section{\label{4}SUMMARY}

In this paper, we have developed the pseudospin representation for the $t$-$J$ model. By introducing pseudospins associated with empty sites, we rewrote the $t$-$J$ model in terms of local spin and spinless fermionic operators. Using this representation, the $t$-$J$ Hamiltonian can be described without any constraint. The pseudospin representation of the $t$-$J$ Hamiltonian was determined by the orientation of the pseudospin state (the parameters $\theta$, $\varphi$). Within a mean-field level, the spin part of the obtained model was regarded as the antiferromagnetic Heisenberg Hamiltonian in a uniform magnetic field with the correspondence $B^{z}/\bar{J}\rightarrow B_{\rm eff}^{z}/J_{\rm eff}\equiv -4t\phi/(J\{(1-\delta)^{2}-\phi^{2}\}+2t\phi)$. The strength and the direction of the effective field are determined by the doping ${\delta}$ and the orientation of pseudospins, respectively. Our method yields reasonable value of the critical doping $\delta_{c}$, at which the antiferromagnetic long-range order disappears, in good agreement with the numerical calculations~\cite{nu1,nu2,nu3,nu4,nu5}. 

\acknowledgments
The authors would like to thank K. Kamide, N. Yokoshi and K. Iigaya for valuable comments and discussions. 
This work is supported by The 21st Century COE Program (Holistic Research and Education center for Physics of Selforganization Systems) at Waseda University from the Ministry of Education, Sports, Culture, Science and Technology of Japan.
\newpage 


\begin{thebibliography}{00} 

\bibitem{tJ1} P. W. Anderson, Science {\bf 235}, 1196 (1987).
\bibitem{tJ2} F. C. Zhang and T. M. Rice, Phys. Rev. B {\bf 37}, 3759 (1988).
\bibitem{tJ3slave5} P. A. Lee, N. Nagaosa, X.-G. Wen, Rev. Mod. Phys. {\bf 78}, 17 (2006) and references therein.
\bibitem{slave1} G. Kotliar and J. Liu, Phys. Rev. B {\bf 38}, 5142 (1988).
\bibitem{slave3} T. Tanamoto, H. Kohno and H. Fukuyama, J. Phys. Soc. Jpn. {\bf 62}, 717 (1993).
\bibitem{slave4} X.-G. Wen and P. A. Lee, Phys. Rev. Lett. {\bf 76}, 503 (1996).
\bibitem{slave6} M. Inui, S. Doniach, P. J. Hirschfeld, and A. E. Ruckenstein, Phys. Rev. B {\bf 37}, 2320 (1988).
\bibitem{slave7} C. L. Kane, P. A. Lee, T. K. Ng, B. Chakraborty, and N. Read, Phys. Rev. B {\bf 41}, 2653 (1990).
\bibitem{slave8} M. Inaba, H. Matsukawa, M. Saitoh, and H. Fukuyama, Physica C {\bf 257}, 299 (1996).
\bibitem{fs1} G. Khaliullin and P. Horsch, Phys. Rev. B {\bf 47}, 463 (1993).
\bibitem{fs2} J. L. Richard and V. Yu. Yushankha\"\i, Phys. Rev. B {\bf 47}, 1103 (1993).
\bibitem{fs3} Y. R. Wang and M. J. Rice, Phys. Rev. B {\bf 49}, 4360 (1994).
\bibitem{fs4} J. Loos, Phys. Rev. B {\bf 53}, 12556 (1996).
\bibitem{fs5} S. \"Ostlund and M. Granath Phys. Rev. Lett. {\bf 96}, 066404 (2006).
\bibitem{mag7} P. Fr\"obrich, P.J. Kuntz, cond-mat/0607675 and references therein.
\bibitem{nu1} T. K. Lee and Shiping Feng, Phys. Rev. B {\bf 38}, 11809 (1988).
\bibitem{nu2} T. Giamarchi and C. Lhuillier, Phys. Rev. B {\bf 43}, 12943 (1991).
\bibitem{nu3} A. Himeda and M. Ogata, Phys. Rev. B {\bf 60}, R9935 (1999).
\bibitem{nu4} M. Ogata and A. Himeda, J. Phys. Soc. Jpn. {\bf 72}, 374 (2003).
\bibitem{nu5} D. A. Ivanov, Phys. Rev. B {\bf 70}, 104503 (2004).
\bibitem{mag1} T. Tamaribuchi and M. Ishikawa, Phys. Rev. B {\bf 43}, 1283 (1991).
\bibitem{mag2} M. E. Zhitomirsky and T. Nikuni, Phys. Rev. B {\bf 57}, 5013 (1998).
\bibitem{mag3} A. Cuccoli, T. Roscilde, V. Tognetti, R. Vaia, and P. Verrucchi, Phys. Rev. B {\bf 67}, 104414 (2003).
\bibitem{mag4} M. E. Zhitomirsky and A. L. Chernyshev, Phys. Rev. Lett. {\bf 82}, 4536 (1999).
\bibitem{mag5} D. Petitgrand, S. V. Maleyev, Ph. Bourges, and A. S. Ivanov, Phys. Rev. B {\bf 59}, 1079 (1999).
\bibitem{mag6} P. J. Jensen and K. H. Bennemann and D. K. Morr and H. Dreysse, Phys. Rev. B {\bf 73}, 144405 (2006).
\bibitem{tyablikov} S. V. Tyablikov, Ukr. Mat. Zh. {\bf 11}, 289 (1959).
\bibitem{tyablikov2} S. V. Tyablikov, {\it Methods in the Quantum Theory of Magnetism} (Plenum Press, New York, 1967).

\end{thebibliography}
%

\end{document}